\documentclass[12pt,runningheads,a4paper]{article}
\usepackage{amsmath}
\usepackage{graphicx}
\usepackage{amssymb}
\usepackage[arrow,matrix]{xy}
\begin{document}
\date{}
\title{Some exact solutions for the rotational flow\\ of a generalized second grade fluid between\\ two circular cylinders}
\maketitle
\author\begin{center}{Amir Mahmood${}^{1\,}\footnote{Corresponding author: Amir Mahmood\\
E-mail address: amir4smsgc@gmail.com\\ Permanent address : Abdus Salam School of Mathematical Sciences, GC
University\\ Lahore, PAKISTAN.}$, Saifullah${}^{1}$, Georgiana Bolat${}^{2}$}
\end{center}
\begin{center}\scriptsize{\textit{ {$^{1}$Abdus Salam School of Mathematical Sciences, GC
University, Lahore, PAKISTAN\\$^{2}$Technical University of Iasi,
R-6600 Iasi, ROMANIA}}}\end{center}
\textbf{Abstract}\\\\{\footnotesize{ \indent The velocity field and
the associated tangential stress corresponding to flow of a
generalized second grade fluid between two infinite coaxial circular
cylinders, are determined by means of the Laplace and Hankel
transforms. At time $t=0$ the fluid is at rest and at $t=0^+$
cylinders suddenly begin to rotate about their common axis with a
constant angular acceleration. The solutions that have been obtained
satisfy the governing differential equations and all imposed initial
and boundary conditions. The similar solutions for a second grade
fluid and Newtonian fluid are recovered from our general solutions.
The influence of the fractional coefficient on the velocity of the
fluid is also analyzed by graphical illustrations.}
\section{Introduction}

\indent \indent A large class of real fluids does not exhibit the linear relationship
between stress and the rate of strain that is now in great interest of scientists and
 engineers. Generally, rheological properties of a material are specified by their so
  called constitutive equations. Among the many constitutive assumptions that have been
  employed to study non-Newtonian fluid behavior, one class that has gained support  from
   both the experimentalists and the theoreticians is that of Rivlin-Ericksen  fluids of
    second grade. The Cauchy stress tensor $\textbf{T}$ for such fluids is given by [1, 2]
\begin{eqnarray}
\textbf{T} = -p\textbf{I} +\mu \textbf{A}_1 + \alpha_1 \textbf{A}_2 + \alpha_2 \textbf{A}_1^2\,,              
\end{eqnarray}
where $p$ is the pressure, $\textbf{I}$ is the unit tensor, $\mu$ is the dynamic viscosity,
 $\alpha_1$ and $\alpha_2$ are the normal stress moduli and $\textbf{A}_1$ and $\textbf{A}_2$
 are the kinematic tensors. In the last years, many authors have made use of rheological equations
 with fractional derivatives to describe the properties of fluids. The constitutive equations with
  fractional derivatives have been proved to be a valuable tool to handle viscoelastic properties.
   In general, these equations are derived from known models by substituting the time ordinary
    derivatives of stress and strain by derivatives of fractional order.

The constitutive equation of the generalized second grade fluids has the same form as (1),
 but the kinematic tensor $\textbf{A}_2$ is defined by [3-5]
\begin{equation}
\textbf{A}_2=D_t^\beta\textbf{A}_1+\textbf{A}_1(\mbox{grad}                 
\textbf{v})+(\mbox{grad}\textbf{v})^T\textbf{A}_1\,,
\end{equation}
where $\textbf{v}$ is the velocity field, $\textbf{A}_1=\mbox{grad}
\textbf{v}+(\mbox{grad}\textbf{v})^T$, the superscript $T$ denotes the transpose operator,
 and $D_t^\beta$ is the Riemann-Liouville fractional derivative operator defined by [4]
\begin{eqnarray}
D^\beta_{t}f(t) =
\frac{1}{\Gamma(1-\beta)}\frac{d}{dt}\int_{0}^{t}\frac{f(\tau)}{(t-\tau)^{\beta}}\,d\tau          
,\,\,\,\,\, \,0<\beta\leq 1\,.
\end{eqnarray}
In the above relation $\Gamma(\cdot)$ is the Gamma function. This model reduces to the
ordinary second grade fluid when $\beta=1$, because $D^1_{t}f=df/dt$.

In this paper, we study the motion of a generalized second grade
fluid between two infinite concentric circular cylinders, both
cylinders are rotating around their common axis $(r=0)$, with
constant angular accelerations. By means of the Laplace and Hankel
transforms we obtain the velocity field and the adequate shear
stress.

\section{Rotational flow between concentric cylinders}

\indent \indent Let us consider an incompressible second grade fluid at rest in an annular region between two straight circular cylinders of radii  $R_1$ and $R_2(>R_1)$. At time $t=0^+$, both cylinders suddenly begin to rotate about their common axis with constant angular accelerations. Owing to the shear, the fluid is gradually moved and its velocity in cylindrical coordinated $(r,\theta,z)$ is given by [2,6]

\begin{eqnarray}
\textbf{v} = \textbf{v} (r,t)= \omega(r,t)\textbf{e}_\theta\,,             
\end{eqnarray}
where  $\textbf{e}_\theta$ is the transverse unit vector.
The basic equations corresponding to this motion are [6,7]

\begin{eqnarray}
\tau(r,t) = (\mu+\alpha_1 D^\beta_{t})(\frac{\partial}            
{\partial r}-\frac{1}{r})\omega(r,t)\,,
\end{eqnarray}

\begin{eqnarray}
\frac{\partial\omega(r, t)}{\partial t}=
(\nu+\alpha D^\beta_{t})\bigg(\frac{\partial^2}
{\partial r^2}+\frac{1}{r}\frac{\partial}{\partial r}-\frac{1}{r^2}\bigg)\omega(r, t),
\,\,\,\,\,\,\,\,r\in(R_1,R_2),\,\,\,t>0,     
\end{eqnarray}
where $\tau(r,t)=S_{r\theta}(r,t)$ is the shear stress which is different of zero, $\nu=\mu/\rho$ is the kinematic viscosity, $\rho$ is the constant density of the fluid and $\alpha= \alpha_1/\rho$.
The appropriate initial and boundary conditions are
\begin{eqnarray}
\omega(r,0)=0\,,                                 
\end{eqnarray}
\begin{eqnarray}
\omega(R_1,t)=R_1\Omega_1t,\,\,\,\,\,\,\,\,\,\,\omega(R_2,t)=R_2\Omega_2t\,\,\,\,\,\mbox{for}\,\,\,\,t>0\,.  
\end{eqnarray}
To solve this problem, we shall use as in [7,8] the Laplace and Hankel transforms.

\subsection{Calculation of the velocity field}
\indent \indent  Applying the Laplace transform to Eqs. (6)-(8) and using the Laplace transform formula for sequential fractional derivatives [4], we obtain the following ordinary differential equation

\begin{eqnarray}
(\nu+\alpha q^\beta)\bigg[\frac{\partial^2\overline{\omega}(r,q)}{\partial r^2}+\frac{1}{r}
\frac{\partial\overline{\omega}(r,q)}{\partial r}-
\frac{\overline{\omega}(r,q)}{ r^2}\bigg]-q\overline{\omega}(r,q)                              
=0\,,
\end{eqnarray}
where the image function $\overline{\omega}(r,q) =\int_0^\infty\omega(r,t)e^{-qt}dt$ \,of\, $\omega(r,t)$ has to satisfy the conditions
\begin{eqnarray}
\overline{\omega}(R_1,q)=\frac{R_1\Omega_1}{q^2}\,,\,\,\,\,\,\,\,\,                        
\overline{\omega}(R_2,q)=\frac{R_2\Omega_2}{q^2}\,,
\end{eqnarray}
$q$ being the transform parameter.

We denote by $\overline{\omega}_{{}_{H}}(r_{n},q)=\int_{R_1}^{R_2}r\overline{\omega}(r,q)B_1(rr_{n})dr$,
the Hankel transform of the function $\overline{\omega}(r,q)$,
where \\$$B_1(rr_{n})=J_1(rr_{n})Y_1(R_2r_{n})-J_1(R_2r_{n})Y_1(rr_{n}),$$\\ and $r_{n}$ are the positive roots of the transcendental equation $B_1(R_1r)=0$ and $J_1(\cdot)$ and $Y_1(\cdot)$ are Bessel functions of order one of the first and second  kind. Applying the Hankel transform to Eq. (9), taking into account the conditions (10) and using the following relations
\begin{eqnarray}
\frac{d}{dr}[B_1(rr_{n})]=r_{n}[J_0(rr_{n})Y_1(R_2r_{n})-J_1(R_2r_{n})Y_0(rr_{n})]            
-\frac{1}{r}B_1(rr_{n})
\end{eqnarray}
and
\begin{eqnarray}
J_0(z)Y_1(z)-J_1(z)Y_0(z)= -\frac{2}{\pi z}\,,           
\end{eqnarray}
we find that
\begin{eqnarray*}
(\nu+\alpha q^\beta)\bigg\{\frac{2[R_2\Omega_2J_1(R_1r_{n})-
R_1\Omega_1J_1(R_2r_{n})]}{\pi
q^2J_1(R_1r_{n})}-r_{n}^2\overline{\omega}_{{}_{H}}(r_{n},q)\bigg\}
-q\overline{\omega}_{{}_{H}}(r_{n},q)=0,
\end{eqnarray*}
or equivalently

\begin{eqnarray}
\overline{\omega}_{{}_{H}}(r_{n},q)=\frac{2[R_2\Omega_2J_1(R_1r_{n})-
R_1\Omega_1J_1(R_2r_{n})]}{\pi J_1(R_1r_{n})}
\frac{\nu+\alpha q^\beta}{q^2[q+\alpha r_{n}^2q^\beta+\nu r_{n}^2]}\,.                      
\end{eqnarray}
Eq. (13) can be written in the following equivalent form

\begin{eqnarray}
\overline{\omega}_{{}_{H}}(r_{n},q)=\overline{\omega}_{{}_{1H}}(r_{n},q)
+\overline{\omega}_{{}_{2H}}(r_{n},q)\,,                                         
\end{eqnarray}
where
\begin{eqnarray}
\overline{\omega}_{{}_{1H}}(r_{n},q)=\frac{2}{\pi r_{n}^2}
\frac{R_2\Omega_2J_1(R_1r_{n})-
R_1\Omega_1J_1(R_2r_{n})}{J_1(R_1r_{n})}\frac{1}{q^2}\,,                       
\end{eqnarray}
and
\begin{eqnarray}
\overline{\omega}_{{}_{2H}}(r_{n},q)=-\frac{2}{\pi r_{n}^2}
\frac{R_2\Omega_2J_1(R_1r_{n})-
R_1\Omega_1J_1(R_2r_{n})}{J_1(R_1r_{n})}
\frac{1}{q[q+\alpha r_{n}^2q^\beta+\nu r_{n}^2]}\,.                       
\end{eqnarray}
The inverse Hankel transforms of the functions
$\overline{\omega}_{{}_{1H}}$ and $\overline{\omega}_{{}_{2H}}$ are
\begin{eqnarray}
\overline{\omega}_{{}_{1}}(r,q)=\frac{\Omega_1 R_1^2(R_2^2-r^2)+
\Omega_2 R_2^2(r^2-R_1^2)}{(R_2^2-R_1^2)r}\,,                                 
\end{eqnarray}
respectively,
\begin{eqnarray}
\overline{\omega}_{{}_{2}}(r,q)=\frac{\pi^2}{2}\sum_{n=1}^\infty
\frac{r_{n}^2J_1^2(R_1r_{n})B_1(rr_{n})}{J_1^2(R_1r_{n})-J_1^2(R_2r_{n})}             
\overline{\omega}_{{}_{2H}}(r_{n},q)\,.
\end{eqnarray}
Now, we find that the function $\overline{\omega}(r,q)$ has the form
\begin{eqnarray*}
\overline{\omega}(r,q)=\frac{\Omega_1 R_1^2(R_2^2-r^2)+
\Omega_2 R_2^2(r^2-R_1^2)}{(R_2^2-R_1^2)r}\frac{1}{q^2}-
\end{eqnarray*}
\begin{eqnarray}
-\pi\sum_{n=1}^\infty
\frac{J_1(R_1r_{n})[R_2\Omega_2J_1(R_1r_{n})-
R_1\Omega_1J_1(R_2r_{n})]B_1(rr_{n})}{J_1^2(R_1r_{n})-J_1^2(R_2r_{n})}
\frac{1}{q[q+\alpha r_{n}^2q^\beta+\nu r_{n}^2]}\,.                                       
\end{eqnarray}
We introduce the notation
\begin{eqnarray}
F(q)=\frac{1}{q[q+\alpha r_{n}^2q^\beta+\nu r_{n}^2]}\,,               
\end{eqnarray}
and rewrite Eq. (20) in the equivalent form
\begin{eqnarray}
F(q)=\frac{q^{-1-\beta}}{(q^{1-\beta}+\alpha r_{n}^2)+\nu r_{n}^2q^{-\beta}}=
\sum_{k=0}^\infty(-\nu r_{n}^2)^k
\frac{q^{-1-\beta-k\beta}}{(q^{1-\beta}+\alpha r_{n}^2)^{k+1}}\,.              
\end{eqnarray}
In order to determine the inverse Laplace transform of the function $\overline{\omega}(r,t)$
we will use the following formulae [9]
\begin{eqnarray*}
L^{-1}\bigg\{\frac{1}{q^a}\bigg\}=\frac{t^{a-1}}{\Gamma(a)}\,,\,\,\,\,\,a>0\,,
\end{eqnarray*}
\begin{eqnarray*}
L^{-1}\bigg\{\frac{q^b}{(q^a-d)^c}\bigg\}=G_{a,b,c}(d,t)=\sum _{j=0}^\infty
\frac{\Gamma(c+j)d^j}{\Gamma(c)\Gamma(j+1)}\frac{t^{(c+j)a-b-1}}{\Gamma[(c+j)a-b]}\,,\,\,\,\,Re(ac-b)>0\,.
\end{eqnarray*}
So we find that the velocity field $\omega(r,t)$ has the following form
\begin{eqnarray*}
\omega(r,t)=\frac{\Omega_1 R_1^2(R_2^2-r^2)+ \Omega_2
R_2^2(r^2-R_1^2)}{(R_2^2-R_1^2)r}t-
\end{eqnarray*}
\begin{eqnarray*}
-\pi\sum_{n=1}^\infty
\frac{J_1(R_1r_{n})[R_2\Omega_2J_1(R_1r_{n})-
R_1\Omega_1J_1(R_2r_{n})]B_1(rr_{n})}{J_1^2(R_1r_{n})-J_1^2(R_2r_{n})}\times
\end{eqnarray*}
\begin{eqnarray}
\times\sum_{k=0}^\infty (-\nu r_{n}^2)^k
G_{1-\beta,-1-\beta-k\beta,k+1} (-\alpha r_{n}^2,t) \,,                               
\end{eqnarray}
or, the equivalently
\begin{eqnarray*}
\omega(r,t)=\frac{\Omega_1 R_1^2(R_2^2-r^2)+ \Omega_2
R_2^2(r^2-R_1^2)}{(R_2^2-R_1^2)r}t-
\end{eqnarray*}
\begin{eqnarray*}
-\pi\sum_{n=1}^\infty
\frac{J_1(R_1r_{n})[R_2\Omega_2J_1(R_1r_{n})-
R_1\Omega_1J_1(R_2r_{n})]B_1(rr_{n})}{J_1^2(R_1r_{n})-J_1^2(R_2r_{n})}\times
\end{eqnarray*}
\begin{eqnarray}
\times\sum_{j,k=0}^\infty \frac{(-\nu r_{n}^2)^k(-\alpha
r_{n}^2)^j\Gamma(k+j+1)}{\Gamma(k+1)\Gamma(j+1)}
\frac{t^{(1-\beta)j+k+1}}{\Gamma[(1-\beta)j+k+2]}\,.                                     
\end{eqnarray}
\subsection{Calculation of the shear stress}
\indent \indent The shear stress $\tau(r,t)$ is obtained from Eqs.
(5) and (23) by means of the Laplace transform.  From Eq. (5) we find
that

\begin{eqnarray}
\overline{\tau}(r,q) = (\mu+\alpha_1 q^\beta)(\frac{\partial}            
{\partial r}-\frac{1}{r})\overline{\omega}(r,q)\,.
\end{eqnarray}
Now, applying the Laplace transform to Eq. (23), differentiating the
result with respect to $r$ and using Eq. (11), we obtain
\begin{eqnarray*}
\overline{\tau}(r,q) =
\frac{2R_1^2R_2^2(\Omega_2-\Omega_1)}{(R_2^2-R_1^2)r^2}(\mu\frac{1}{q^2}+\alpha_1\frac{1}{q^{2-\beta}})
+\pi\sum_{n=1}^\infty\bigg[\frac{2}{r}B_1(rr_{n})-r_{n}B(rr_{n})\bigg]\times
\end{eqnarray*}
\begin{eqnarray*}
\times\frac{J_1(R_1r_{n})[R_2\Omega_2J_1(R_1r_{n})-
R_1\Omega_1J_1(R_2r_{n})]}{J_1^2(R_1r_{n})-J_1^2(R_2r_{n})}
\sum_{j,k=0}^\infty\frac{(-\nu r_{n}^2)^k(-\alpha
r_{n}^2)^j\Gamma(k+j+1)}{\Gamma(k+1)\Gamma(j+1)}\times
\end{eqnarray*}
\begin{eqnarray}
\times\bigg[\mu\frac{1}{q^{(1-\beta)j+k+2}}+\alpha_1\frac{1}{q^{(1-\beta)j+k+2-\beta}}\bigg]\,,    
\end{eqnarray}
where
\begin{eqnarray*}
B(rr_{n})=J_0(rr_{n})Y_1(R_2r_{n})-J_1(R_2r_{n})Y_0(rr_{n})\,.
\end{eqnarray*}
Applying inverse Laplace transform to the image function
$\overline{\tau}(r,q)$, we find the shear stress
\begin{eqnarray*}
\tau(r,t) =
\frac{2R_1^2R_2^2(\Omega_2-\Omega_1)}{(R_2^2-R_1^2)r^2}(\mu
t+\frac{\alpha_1t^{1-\beta}}{\Gamma(2-\beta)})
+\pi\sum_{n=1}^\infty\bigg[\frac{2}{r}B_1(rr_{n})-r_{n}B(rr_{n})\bigg]\times
\end{eqnarray*}
\begin{eqnarray*}
\times\frac{J_1(R_1r_{n})[R_2\Omega_2J_1(R_1r_{n})-
R_1\Omega_1J_1(R_2r_{n})]}{J_1^2(R_1r_{n})-J_1^2(R_2r_{n})}
\sum_{j,k=0}^\infty\frac{(-\nu r_{n}^2)^k(-\alpha
r_{n}^2)^j\Gamma(k+j+1)}{\Gamma(k+1)\Gamma(j+1)}\times
\end{eqnarray*}
\begin{eqnarray}
\times\bigg[\mu\frac{t^{(1-\beta)j+k+1}}{\Gamma[(1-\beta)j+k+2]}+
\alpha_1\frac{t^{(1-\beta)j+k+1-\beta}}{\Gamma[(1-\beta)j+k+2-\beta]}\bigg]\,.                
\end{eqnarray}
\section{Limiting case $(\beta=1)$}
\indent \indent  Making $\beta=1$ into Eq. (22) we obtain the
velocity field
\begin{eqnarray*}
\omega(r,t)=\frac{\Omega_1 R_1^2(R_2^2-r^2)+ \Omega_2
R_2^2(r^2-R_1^2)}{(R_2^2-R_1^2)r}t-
\end{eqnarray*}
\begin{eqnarray*}
-\pi\sum_{n=1}^\infty
\frac{J_1(R_1r_{n})[R_2\Omega_2J_1(R_1r_{n})-
R_1\Omega_1J_1(R_2r_{n})]B_1(R_1r_{n})}{J_1^2(R_1r_{n})-J_1^2(R_2r_{n})}\times
\end{eqnarray*}
\begin{eqnarray}
\times\sum_{k=0}^\infty(-\nu r_{n}^2)^k
G_{0,-2-k,k+1}(-\alpha r_{n}^2,t)\,,                                      
\end{eqnarray}
corresponding to an ordinary second grade fluid, performing the same motion.
Similarly, from (26), we obtain the associated shear stress

\begin{eqnarray*}
\tau(r,t) =
\frac{2R_1^2R_2^2(\Omega_2-\Omega_1)}{(R_2^2-R_1^2)r^2}(\mu
t+\alpha_1)
+\pi\sum_{n=1}^\infty\bigg[\frac{2}{r}B_1(rr_{n})-r_{n}B(rr_{n})\bigg]\times
\end{eqnarray*}
\begin{eqnarray*}
\times\frac{J_1(R_1r_{n})[R_2\Omega_2J_1(R_1r_{n})-
R_1\Omega_1J_1(R_2r_{n})]}{J_1^2(R_1r_{n})-J_1^2(R_2r_{n})}
\sum_{j,k=0}^\infty\frac{(-\nu r_{n}^2)^k(-\alpha
r_{n}^2)^j\Gamma(k+j+1)}{\Gamma(k+1)\Gamma(j+1)}\times
\end{eqnarray*}
\begin{eqnarray}
\times\bigg[\mu\frac{t^{k+1}}{\Gamma(k+2)}+
\alpha_1\frac{t^{k}}{\Gamma(k+1)}\bigg]\,,                                                      
\end{eqnarray}
The above relations can be simplified if we use the following relations:

\begin{eqnarray*}
\sum_{k=0}^\infty(-\nu r_{n}^2)^k
G_{0,-2-k,k+1}(-\alpha r_{n}^2,t)=\sum_{k=0}^\infty(-\nu r_{n}^2)^k
\sum_{j=0}^\infty\frac{(-\alpha
r_{n}^2)^j\Gamma(k+j+1)}{\Gamma(k+1)\Gamma(j+1)}\frac{t^{k+1}}{\Gamma(k+2)}=
\end{eqnarray*}

\begin{eqnarray*}
=\sum_{k=0}^\infty\frac{(-\nu r_{n}^2)^kt^{k+1}}{\Gamma(k+2)}
\frac{1}{(1+\alpha r_{n}^2)^{k+1}}=-\frac{1}{\nu r_{n}^2}\sum_{k=0}^\infty
\frac{1}{(k+1)!}\bigg(-\frac{\nu r_{n}^2t}{1+\alpha r_{n}^2}\bigg)^{k+1}=
\end{eqnarray*}
\begin{eqnarray*}
=\frac{1}{\nu r_{n}^2}\bigg[1-\exp\bigg(-\frac{\nu r_{n}^2t}{1+\alpha r_{n}^2}\bigg)\bigg]\,.
\end{eqnarray*}
As a result, we find that, the velocity field has the form

\begin{eqnarray*}
\omega(r,t)=\frac{\Omega_1 R_1^2(R_2^2-r^2)+ \Omega_2
R_2^2(r^2-R_1^2)}{(R_2^2-R_1^2)r}t-
\end{eqnarray*}
\begin{eqnarray}
-\frac{\pi}{\nu}\sum_{n=1}^\infty
\frac{J_1(R_1r_{n})[R_2\Omega_2J_1(R_1r_{n})-
R_1\Omega_1J_1(R_2r_{n})]}{J_1^2(R_1r_{n})-J_1^2(R_2r_{n})}\frac{B_1(rr_{n})}{r_{n}^2}
\bigg[1-\exp\bigg(-\frac{\nu r_{n}^2t}{1+\alpha r_{n}^2}\bigg)\bigg]\,,                       
\end{eqnarray}
and the shear stress has the form
\begin{eqnarray*}
\tau(r,t) =
\frac{2R_1^2R_2^2(\Omega_2-\Omega_1)}{(R_2^2-R_1^2)r^2}(\mu
t+\alpha_1)
+\pi\sum_{n=1}^\infty\bigg[\frac{2}{r}B_1(rr_{n})-r_{n}B(rr_{n})\bigg]\times
\end{eqnarray*}
\begin{eqnarray*}
\times\frac{J_1(R_1r_{n})[R_2\Omega_2J_1(R_1r_{n})-
R_1\Omega_1J_1(R_2r_{n})]}{J_1^2(R_1r_{n})-J_1^2(R_2r_{n})}
\bigg\{\frac{\mu}{\nu r_{n}^2}
\bigg[1-\exp\bigg(-\frac{\nu r_{n}^2t}{1+\alpha r_{n}^2}\bigg)\bigg]+                 
\end{eqnarray*}
\begin{eqnarray}
+\frac{\alpha_1}{1+\alpha r_{n}^2}
\exp\bigg(-\frac{\nu r_{n}^2t}{1+\alpha r_{n}^2}\bigg)\bigg\}\,.
\end{eqnarray}
Eqs. (29) and (30) are identical with those obtained by Fetecau
\emph{et al} [6 , Eqs. (3.12) and (3.16) for $\lambda\rightarrow0$].
Making $\alpha\rightarrow0$ into Eqs. (29) and (30), the similar
solutions corresponding to the Newtonian fluid, performing the same
motion, are recovered. Making $\Omega_1=0$ and $\Omega_2=\Omega$ or
$\Omega_1=\Omega$ and $\Omega_2=0$ into Eqs. (23) and (26), we
obtain the velocity field and the adequate shear stress
corresponding to the flow between two cylinders, one of them being
at rest.
\section{Conclusion and numerical results}
In this paper we establish exact solutions for the velocity field and shear stress corresponding
 to the flow of a generalized second grade fluid between two concentric circular cylinders.
 The motion is produced by the two cylinders which at time $t=0^+$ begin to rotate around their
  common axis with angular velocities $\Omega_1t$ and $\Omega_2t$. The solutions, obtained by means
  of Laplace and Hankel transforms, are presented under integral and series forms in terms of the
   generalized $G$-function, and satisfy all imposed initial and boundary conditions. For $\beta=1$
   or $\beta=1$ and $\alpha=0$, the similar solutions for the ordinary second grade fluids, respectively,
    Newtonian fluids are recovered. The velocity field and adequate shear stress corresponding to the flow
     between two cylinders, one of them being at rest, are obtained as particular cases of our general
     solutions. Making $\Omega_1=0$ and $\Omega_2=\Omega$ into Eqs. (23), for instance, we obtain the velocity field
\begin{eqnarray*}
\omega(r,t)=\frac{\Omega
R_2^2(r^2-R_1^2)}{(R_2^2-R_1^2)r}t-\pi R_2\Omega\sum_{n=1}^\infty
\frac{J_1^2(R_1r_{n})B_1(rr_{n})}{J_1^2(R_1r_{n})-J_1^2(R_2r_{n})}\times
\end{eqnarray*}
\begin{eqnarray}
\times\sum_{j,k=0}^\infty \frac{(-\nu r_{n}^2)^k(-\alpha
r_{n}^2)^j\Gamma(k+j+1)}{\Gamma(k+1)\Gamma(j+1)}
\frac{t^{(1-\beta)j+k+1}}{\Gamma[(1-\beta)j+k+2]}\,,                                     
\end{eqnarray}
corresponding to the flow between cylinders, the inner cylinder being at rest.

Finally, the numerical results are given to illustrate the influence of the fractional parameter $\beta$ on the velocity $\omega(r,t)$. In all figures we considered $R_1=1$, $R_2=4$, $\Omega_1=3$, $\Omega_2=1.5$, $\rho=1260$, $\alpha_1=11.34$ and $\mu=1.48$.

In Figs. 1 the profiles of the velocity $\omega(r,t)$, corresponding to the motion of Newtonian fluid (the curve $\omega\mbox{N}(r)$), second grade fluid (the curve $\omega\mbox{SG}(r)$)) and generalized second grade fluid (the curves $\omega1(r), \omega2(r)\,\,\,\mbox{and}\,\,\, \omega3(r)$), are plotted, for different values of the fractional coefficient $\beta$ and time $t$. It is clearly seen from these figures that velocity increases when the fractional coefficient decreases. Moreover, the influence of $\beta$ is more strong near boundary of the domain and the generalized second grade fluid flows faster than the second grade and Newtonian fluids.

Figs. 2 depict the histories of the velocity field $\omega(r,t)$ at the positions $r=1.3,\, 2.5\,\,\,\mbox{and}\,\,\,3.8$, for $t\in[0,10]$ and different values of $\beta$. One can see that the influence of $\beta$ is more strong near boundary of the domain and the velocity increases when $\beta$ decreases. The units of the parameter in Figs. 1-2 are from SI units and the roots $r_n$ have been approximated by $n\pi/(R_2-R_1)$ [10].

\newpage

\begin{figure}
  \includegraphics[width=8cm]{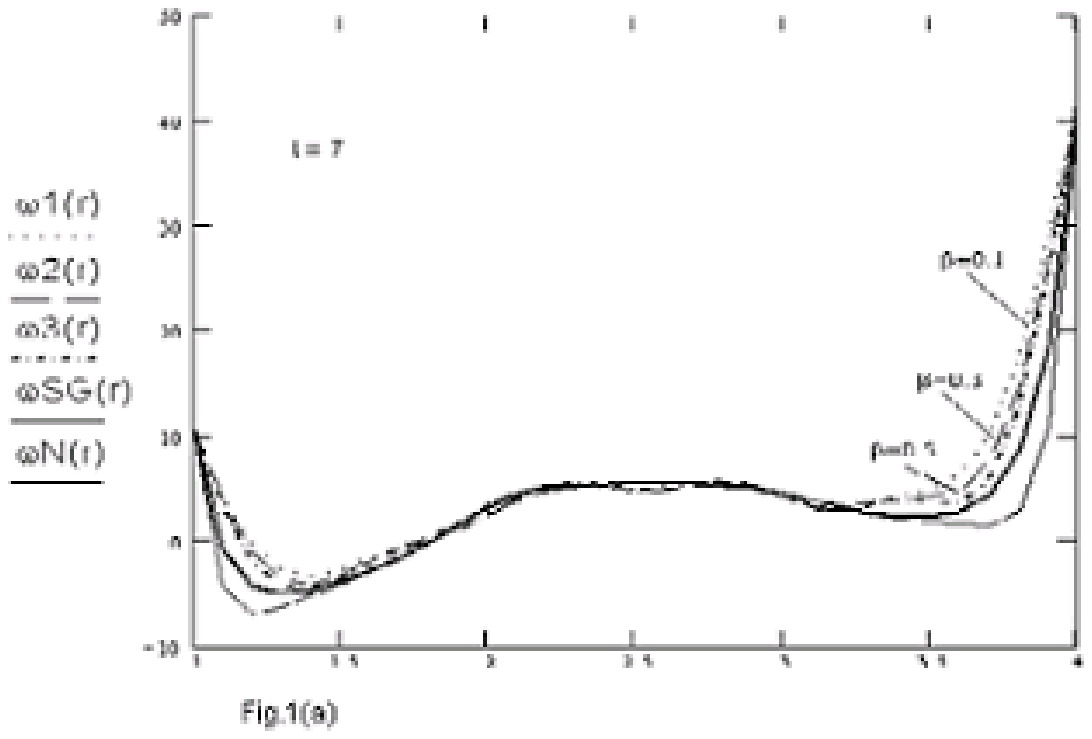}
\end{figure}
\begin{figure}
  \includegraphics[width=8cm]{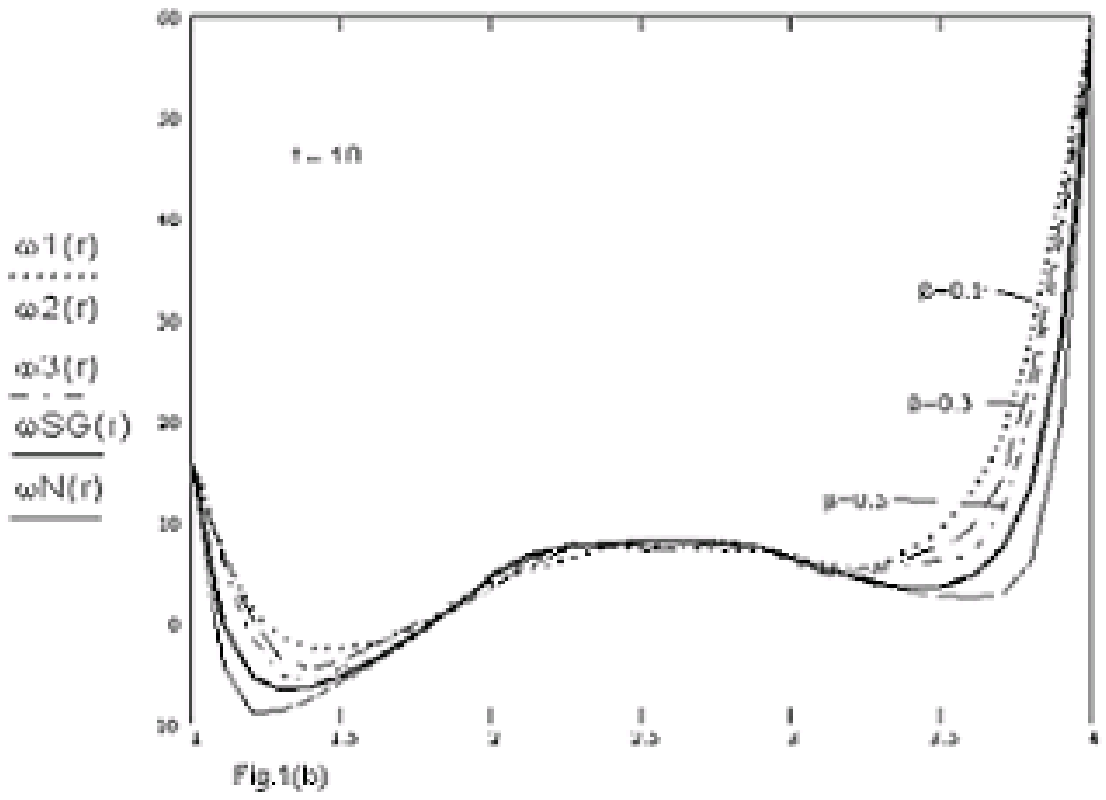}
\end{figure}
\begin{figure}
  \includegraphics[width=8cm]{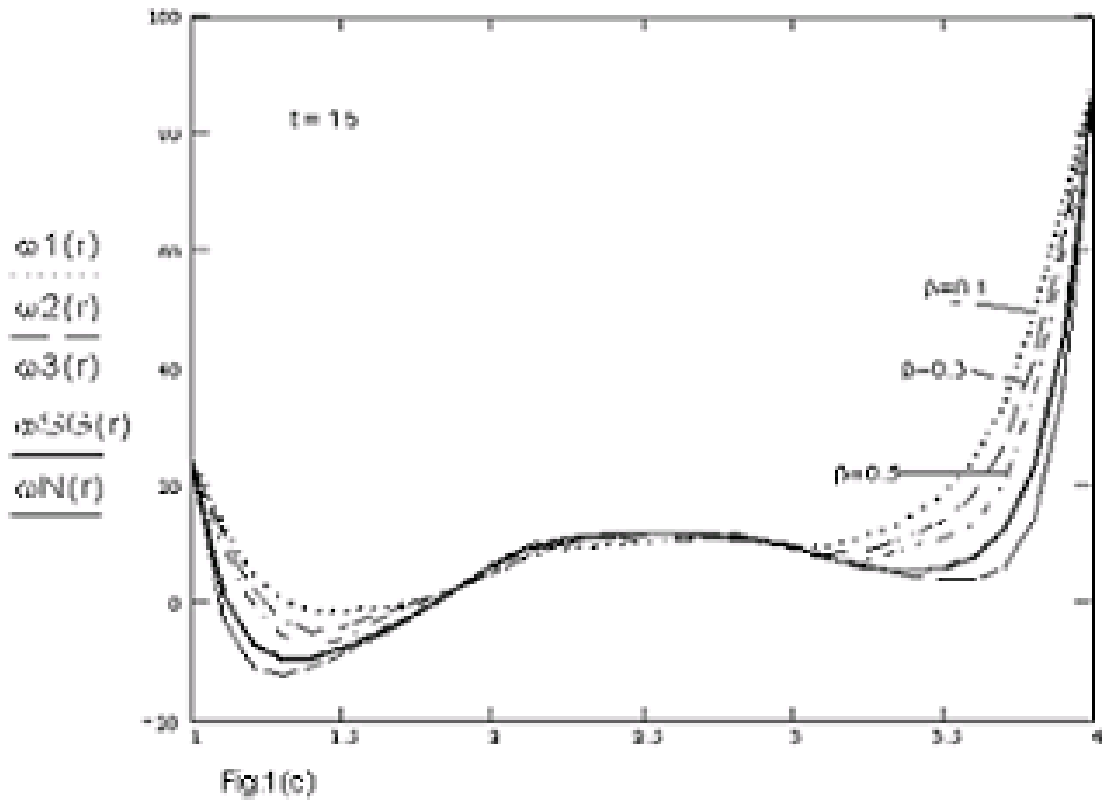}
  \caption{Velocity profile $\omega(r)$ for different values of the fractional coefficient $\beta$}\label{1}
\end{figure}
\begin{figure}
  \includegraphics[width=8cm]{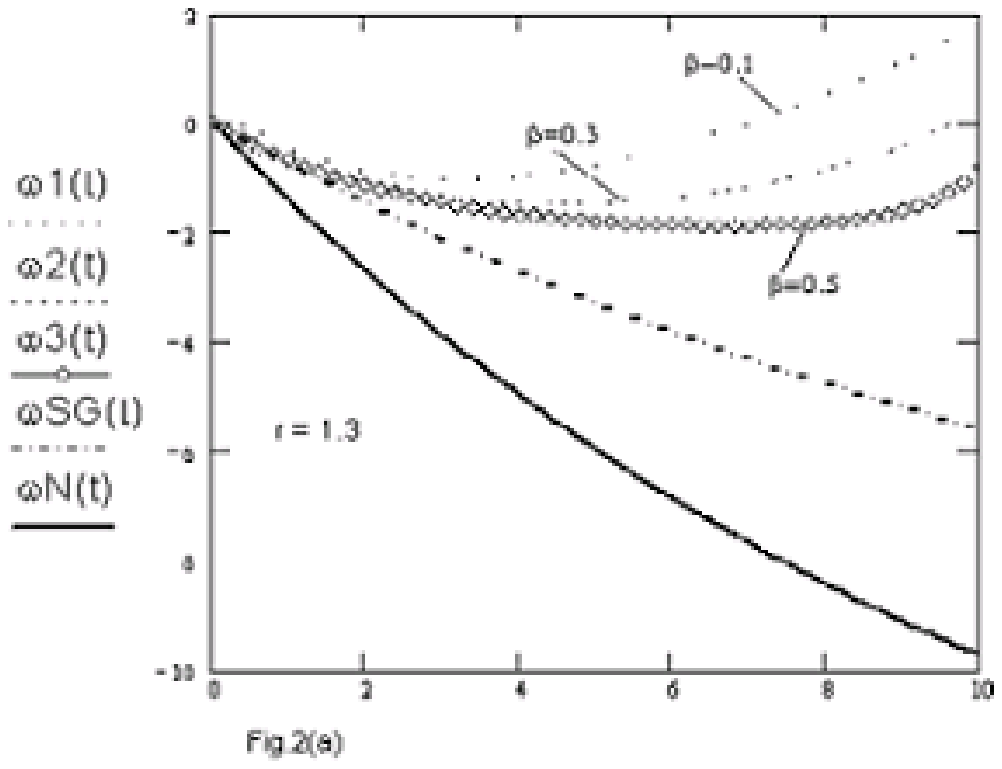}
\end{figure}
\begin{figure}
  \includegraphics[width=8cm]{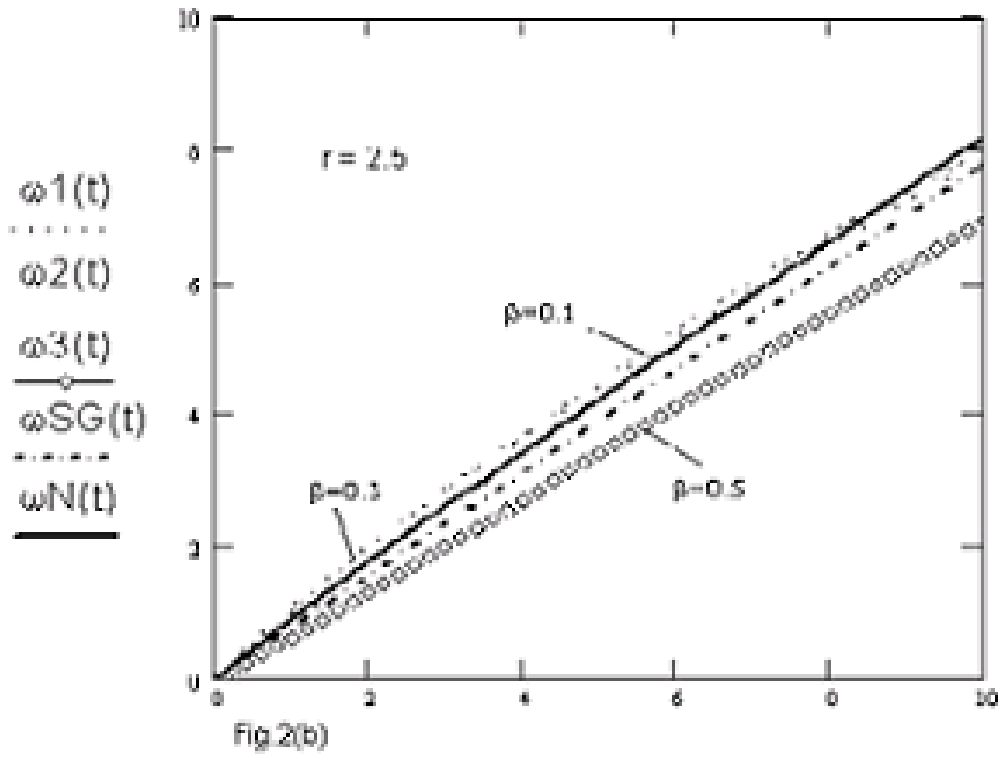}
\end{figure}
\begin{figure}
  \includegraphics[width=8cm]{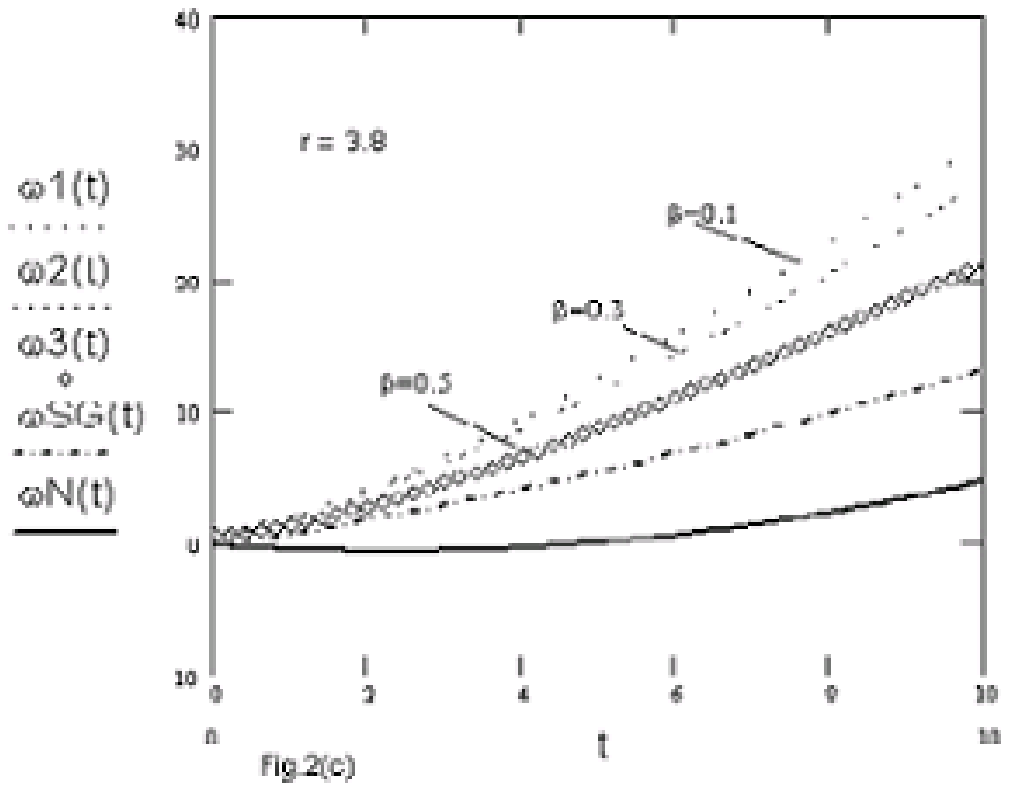}
  \caption{Time variation of the velocity}\label{1}
\end{figure}

\end{document}